\documentstyle[prb,aps,multicol,eqsecnum,epsf]{revtex}
\begin{document}
\title{Unusual Properties of Hall Gas: 
Implication to Metrology of the Integer Quantum Hall Effect} 
\author{K. Ishikawa and N. Maeda}
\address{
Department of Physics, Hokkaido University, 
Sapporo 060-0810, Japan}
\maketitle
\begin{abstract}
Physical properties of anisotropic compressible quantum Hall states 
and their implications to integer quantum Hall effect are studied based on  
a mean field theory on the von Neumann lattice. 
It is found that the Hall gas has unusual thermodynamic properties such as 
negative pressure and negative compressibility and unusual transport 
properties. 
Transport properties and density profile of Hall gas states at half fillings
agree with those of anisotropic states discovered experimentally in
higher Landau levels. 
Hall gas formed in the bulk does not spread but shrinks, owing to negative 
pressure, and a strip of Hall gas gives abnormal electric transport at finite 
temperature.
Conductances at finite temperature and finite injected current agree with 
recent experiments on collapse and breakdown phenomena of the integer quantum 
Hall effect. 
As a byproduct, existence of  new quantum Hall regime, dissipative quantum 
Hall regime, in which Hall resistance is 
quantized exactly even in the system of small longitudinal resistance is 
derived. 
\end{abstract}
\draft

\begin{multicols}{2}

\section{Introduction}
In the standard view of the integer quantum Hall effects 
(IQHE),\cite{a,metro} the Hall resistance 
is quantized as $h\over e^2 n$ when the longitudinal resistance vanishes. 
This is caused by random disorders. 
Localization effects due to disorders are 
peculiar in two-dimensional electron system with strong magnetic field, 
which we call the quantum Hall system hereafter. 
Level density which is composed of discrete 
and degenerate Landau levels in the system without disorders becomes to be 
composed of non-degenerate discrete levels in addition to degenerate levels 
in the system of disorders. 
The one-particle states in the center of energetically broadened Landau 
levels are extended and 
those states outside the energy regions are localized. 
Localized states have 
discrete energies and do not carry the bulk electric current. 
Hence if the Fermi energy is in the localized states region, 
the longitudinal resistance vanishes 
and there is no energy dissipation. 
The Hall current is carried by the electrons in the extended states 
below the Fermi energy and the Hall resistance is quantized. 
Thus the Hall resistance is quantized when the longitudinal resistance 
vanishes. 
Conversely if the Fermi energy is in the extended state regions and electrons
in these extended states would behave like 
ordinary gas and spread over whole spatial region, these electrons in the 
extended states cover whole spatial region and carry bulk electric current. 
Hall resistance and longitudinal 
resistance are determined by current correlation functions and extended states
give finite contributions. 
Consequently the Hall resistance deviates from 
one of the quantized value by an amount which is proportional to the 
longitudinal resistance. 

In the finite system there are edges and current contacts. 
They play important roles in IQHE of finite systems. 
In the system of finite injected current, the potential 
around the current contacts 
is quite singular, because Hall electric field is finite in the quantum Hall 
system but vanishes in the inside of current contacts. 
There is a finite 
potential drop along the injected current. 
Hence there is always energy 
dissipation around the current contacts and the bulk Hall resistance becomes 
unquantized.
The physical phenomena 
in other region, if they are caused by electrons in the extended states, are
causally connected with those of current contact regions. 
Thus if there are 
extended states at the Fermi energy, Hall resistance measured at any regions 
is unquantized. 
It is impossible to have the spatial region where there 
is no energy dissipation and the Hall resistance is quantized. 
The Hall resistance would never been quantized, then, 
regardless of the position where the Hall potential is measured. 
On the other hand, if the localized electrons 
causes the singular behavior around the current contact, the 
effect is confined in this region and does not give any effects at other 
regions. 
The Hall resistance measured at different but suitable positions can 
be quantized. 
Whether there exist regions where the Hall resistance is quantized, it is 
important for the electrons at the Fermi energy to be localized or to be 
confined in finite region. 
Hence the thermodynamic properties of Hall gas is 
important for the metrology of IQHE. 

Actually unusual properties such as negative compressibility of Hall gas have 
been known. 
We use the terminology, Hall gas, for a many-body state of having 
continuous one-particle energy in the quantum Hall system in this paper. 
Several puzzling phenomena also have been observed not only in the 
medium mobility samples but also in the high mobility samples.  
In the medium mobility samples, the quantization of the Hall resistance 
sometimes coexists with small longitudinal resistance and unusual breakdown 
phenomena of IQHE are induced by injected current. 
In the latter, IQHE are broken in two steps \cite{b}. 
The Hall resistance becomes to deviate from the quantized value in the first 
step and the longitudinal resistance becomes large value in the second step. 
The first step is named the collapse of IQHE and the second step is named 
the breakdown of IQHE.\cite{ebert,cage,komiyama,kawa} 
In high mobility sample, 
highly anisotropic compressible states are discovered recently around half 
filling at higher Landau levels.\cite{c,d,e,f} 
Hartree-Fock calculations\cite{i,j,k} and other numerical 
calculations\cite{HAL,DMRG} on the stripe state 
seem to agree with the experiments. 
Although this state is called the stripe state or unidirectional charge 
density wave from its density profile, 
we prefer to use anisotropic Hall gas. 
This is because in this state Fermi surface is defined. 
Physical properties of the many-electron state is 
determined mainly from the electrons near Fermi surface. 
So it is quite valuable to identify the Fermi surface. 
It is seen clearly that the present Hartree-Fock states has anisotropic 
Fermi surface and this many-body state behaves like gas. 
Suitable representation that has a translational invariance
is convenient to see this behavior explicitly. 
The von Neumann lattice representation is such representation that 
meets this requirement. 
We use this  representation and study physical 
properties of the Hall gas state. 
We show that most of physical properties become clear from our viewpoint, 
which is the advantage of our method. 
Although  it was predicted that fluctuation effects change the stripe state 
into the liquid crystal or stripe crystal\cite{fra,macf,co}, 
we regard fluctuation effects are irrelevant to physics 
found experimentally in higher Landau levels, 
as long as the anisotropic property is kept. 
We estimate higher order corrections of the energy and others from the 
residual interaction and find actually that the corrections are small. 
Some physical properties have been given already in Ref.(\cite{imo} for the 
anisotropic compressible state and a connection with metrology of IQHE has 
been pointed out partly \cite{pb}. 
We give a full argument in this work.

We show in the present work that the Hall gas has unusual thermodynamic 
properties and that the above puzzling phenomena become understandable. 
The unusual properties of Hall gas insure the stability of IQHE. 
Furthermore a new quantum Hall regime with energy dissipations is derived.

Hall gas of anisotropic Fermi surface obtained based on mean field theory on 
von Neumann lattice is studied first. 
The Hamiltonian has a translational 
invariance in the momentum space, called K-symmetry.\cite{ks} 
The symmetry is 
broken spontaneously in the Hall gas in which a single particle energy is 
generated self-consistently \cite{imo}. 
This Hall gas with uniform neutralizing
background charge has negative pressure and compressibility. 
The Hall gas state which preserves $K_x$-symmetry and breaks 
$K_y$-symmetry has Fermi surface parallel to $p_x$ axis. 
In $p_x$ direction whole one-particle states within one Landau level are 
filled completely and lowest energy empty states are in the next Landau 
level. 
There is an energy gap of Landau level in $p_x$-direction but there is no 
energy gap in the perpendicular direction. 
Longitudinal resistance in this x-direction vanishes but the longitudinal 
resistance orthogonal to this direction becomes finite. 

Because center coordinates of Hall gas are non-commuting,
the density profile of $K_x$ invariant anisotropic 
state is uniform in y-direction but is periodic in x-direction.\cite{k} 
Thus Fermi sea in 
momentum space and density in coordinate space become orthogonal.      
The system has anisotropic resistances and is 
regarded as anisotropic compressible gas   
observed at higher Landau levels of high mobility samples.\cite{c,d,e,f} 
From its density profile the present anisotropic compressible state is 
equivalent to the stripe state\cite{g,h} or unidirectional charge 
density wave.\cite{kura}
Response of the system under a periodic density modulation will be 
discussed also.

In medium mobility samples, disorders are important. 
In the present work, long range disorder potential 
which is realized in GaAs samples is studied. 
Due to interplay between finite size effects and 
long range disorder effects a strip of conducting electron states 
is generated. 
Compressible states formed by interaction in the localized state energy 
region, where one-particle energy is discrete, have negative compressibility.  
\cite{efros,eisen} 
The strip is then formed. 
If the system has a finite size, furthermore, strip 
appears even without interaction from finite size effect. 
Localized electrons with localization lengths of the system size or larger 
behave like the extended states. 
If the localization lengths are larger than the width but shorter than 
the length of the system, this state in the bulk region is like strip 
and is disconnected with either the source or drain region. 

We show unusual transport of the isolated strip in the bulk next. 
Strip of extended electrons at Fermi energy contributes to electric 
resistances through tunneling processes. 
If the injected current flows through the extended states in the lower 
Landau levels, the current is induced in the strip of extended states at Fermi
energy at a finite temperature by tunneling. 
Magnitude of the current and resistance are computed. 
In the Hall bar sample with suitable sizes, the Hall 
resistance can be quantized with a finite longitudinal resistance.  
This is a new regime of the IQHE and could occur at the finite injected 
current near the localized state region, i.e., at $\nu=n+\delta$. 
The strip thus leads to a new regime of IQHE, i.e., dissipative quantum Hall 
regime at $\nu=n+\delta$. 
The unusual IQHE and breakdown due to current are also derived. 
Thus the unusual properties of the Hall gas states naturally 
explain unusual IQHE and breakdown observed in experiments. 

The present paper is organized in the following way. 
In Sec.~II, the anisotropic compressible mean field solutions 
obtained before is studied and their thermodynamic properties and the 
responses of the system to external potential modulation are computed. 
It is shown that anisotropic Hall gas has negative 
pressure and compressibility. 
In Sec.~III, strip of the Hall gas states is analyzed. 
Owing to the negative pressure, compressible gas states which are 
composed of electrons in extended states shrink and form strips. 
Consequently they do not reach source or drain regions and do not contribute 
to the bulk electronic conductances at zero temperature. 
In Sec.~IV, transport properties of anisotropic compressible 
states at zero temperature are analyzed and compared with the recent 
experiments. 
Current activation from lower Landau levels to the strip of compressible gas 
around Fermi energy at finite temperature is analyzed in Sec.~V 
and summary is given in Sec.~VI. 

\section{Compressible mean field states: thermodynamic properties}
\subsection{Von Neumann lattice representation}

We use the von Neumann lattice representation,\cite{l,m} which treats 
two-dimensional spaces symmetrically and preserves two-dimensional 
translational invariance. 
Hence, electron operators have momentum, $\bf p$, 
which is defined in the magnetic Brillouin zone and Landau level index, $l$. 

The one-body Hamiltonian,
\begin{equation}
{\hat H}_0={1\over2m}({\bf p}+e{\bf A})^2,\ 
\partial_x A_y-\partial_y A_x=B,
\label{one}
\end{equation}
is written by the two sets of coordinates, 
guiding center coordinates $(X,Y)$ and relative coordinates $(\xi,\eta)$ 
as, 
\begin{eqnarray}
&{\hat H}_0={m\omega^2_c\over2}(\xi^2+\eta^2),\ \omega_c={eB\over m},
\nonumber\\
&X=x-\xi,\ Y=y-\eta,\\
&\xi=(eB)^{-1}(p_y+eA_y),\ \eta=-(eB)^{-1}(p_x+eA_x).
\nonumber
\end{eqnarray}
These coordinates satisfy commutation relations,
\begin{eqnarray}
&[\xi,\eta]=-[X,Y]={a^2\over2\pi i},\label{comm}\\
&[\xi,X]=[\xi,Y]=[\eta,X]=[\eta,Y]=0,
\nonumber
\end{eqnarray}
where $a=\sqrt{2\pi\hbar/eB}$. 
Since the two sets of coordinates are commutative, the Hilbert space 
is spanned by the direct product of two spaces on which each sets of 
coordinates operate. 

A discrete set of coherent states of guiding center variables, 
\begin{eqnarray}
&(X+iY)\vert\alpha_{mn}\rangle=z_{mn}\vert\alpha_{mn}\rangle,\nonumber\\
&z_{mn}=a(mr+i{n\over r}),\ m,\ n;{\rm\ integers},\\
&\vert\alpha_{mn}\rangle=(-1)^{m+n+mn}e^{\sqrt{\pi}(A^\dagger 
{z_{mn}\over a}-A {z_{mn}^*\over a})}\vert\alpha_{00}\rangle,
\nonumber\\ 
&A={\sqrt{\pi}\over a}(X+iY), 
\nonumber
\end{eqnarray}
is a complete set. 
These coherence states are localized at the lattice $a(mr,n/r)$, 
where a positive real number $r$ is the asymmetric parameter of the unit cell. 
Matrix elements of two different coherent states are 
translational invariant and the Fourier transformed states satisfy,
\begin{eqnarray}
&\vert\alpha_{\bf p}\rangle=\sum_{mn} e^{ip_x m+ip_y n}
\vert\alpha_{mn}\rangle,\nonumber\\
&\langle\alpha_{\bf p}\vert\alpha_{\bf p'}\rangle=(2\pi)^2\sum_N 
\delta({\bf p}-{\bf p'}-2\pi {\bf N})\beta({\bf p})\beta^*({\bf p}),\\
&\beta({\bf p})=(2{\rm Im}\tau)^{1/4}e^{{i\tau p_y^2\over4\pi}}
\Theta_1({p_x+\tau p_y\over 2\pi}\vert\tau),\nonumber
\end{eqnarray}
where $\Theta_1$ is a Jacobi's theta function and $\tau=ir^2$. 
The theta function vanishes at ${\bf p}=0$ and the corresponding function 
has zero norm. 
The states of finite momentum have finite norms and two states of different 
momentum are orthogonal. 
Orthonormalized states are defined as 
\begin{equation}
\vert\beta_{\bf p}\rangle=
{1\over\beta({\bf p})e^{i\chi({\bf p})}}\vert\alpha_{\bf p}\rangle.
\end{equation}
The real function $\chi$ represents the gauge degree of freedom in the 
momentum space. 
The eigenfunctions of one-body free Hamiltonian, Eq.~(\ref{one}), satisfy, 
\begin{equation}
{m\omega^2_c\over2}(\xi^2+\eta^2)\vert f_l\rangle=
\hbar\omega_c(l+{1\over2})\vert f_l\rangle.
\end{equation}
The Hilbert space is spanned by the direct product of the eigenfunctions 
\begin{equation}
\vert l,{\bf p}\rangle=\vert f_l\rangle\otimes\vert\beta_{\bf p}\rangle.
\end{equation}
The explicit form in the real space of the eigenfunction, $\langle {\bf x}
\vert l,{\bf p}\rangle$, is given in Ref.~(\cite{m}). 

\subsection{Hamiltonian of interacting two-dimensional electrons in the 
quantum Hall system and its symmetry}
We expand the electron field, 
\begin{equation}
\Psi({\bf x})=\int_{\rm BZ}{d^2p\over(2\pi)^2}\sum_{l=0}^{\infty} 
b_l({\bf p})\langle {\bf x}\vert l,{\bf p}\rangle.
\label{ele}
\end{equation}
Electrons in quantum Hall systems are described by the expansion coefficients 
that are specified by momentum $\bf p$ and Landau level index $l$. 
We substitute  this expansion into the action and develop and study  
field theory of the quantum Hall dynamics.
Energy, thermodynamic properties, 
and energy corrections due to small density modulations of compressible Hall 
gas are computed in this section. 

The electron operators satisfy the commutation relation,
\begin{eqnarray}
&\{b_l({\bf p}),b^\dagger_{l'}({\bf p'})\}\delta(t-t')=\qquad\qquad\qquad
\qquad\qquad\nonumber\\
&\quad\delta_{ll'}\sum_{\bf N}(2\pi)^2\delta({\bf p}-{\bf p'}-2\pi{\bf N})
e^{i\phi({\bf p,N})}\delta(t-t'),\\
&b_l({\bf p}+2\pi{\bf N})=e^{i\phi({\bf p,N})}b_l({\bf p}),
\nonumber\\
&\phi({\bf p,N})=\pi(N_x+N_y)-N_y p_x.
\nonumber
\end{eqnarray}
The total Hamiltonian of charge neutral system in which background charge 
cancels the electron's charge is given by,
\begin{eqnarray}
&H=H_0+H_1,\nonumber\\
&H_0=\sum_l\int_{\rm BZ}
{d^2p\over(2\pi)^2}E_l b_l^\dagger({\bf p})b_l({\bf p}),
\label{hami}\\
&H_1=\int_{{\bf k}\neq0}{d^2k\over(2\pi)^2}
\rho({\bf k}){{\tilde V}(k)\over2}
\rho(-{\bf k}),\nonumber
\end{eqnarray}
with the electron density operator
\begin{eqnarray}
\rho({\bf k})&=&\int_{\rm BZ}{d^2p\over(2\pi)^2}
b^\dagger_l({\bf p})b_{l'}({\bf p}-a\hat{\bf k})f_{ll'}({\bf k})
\label{ham}\\
&&\times\exp[-i{a\over4\pi}{\hat k}_x (2p_y-a {\hat k}_y)],\nonumber\\
f_{ll'}({\bf k})&=&\langle f_l\vert e^{ik\cdot\xi}\vert f_{l'}\rangle,
\nonumber
\end{eqnarray}
where $\hat k=(r k_x,k_y/r)$, ${\tilde V}(k)=2\pi q^2/k$ ($q^2=e^2/\epsilon$, 
$\epsilon$ is a dielectric constant). 
The explicit form of $f_{ll'}$ is written by the Laguerre polynomial 
and is given in Ref(\cite{m}). 
In particular, $f_{ll}({\bf k})=L_l(a^2k^2/4\pi)e^{-a^2 k^2/8\pi}$. 
This Hamiltonian has peculiar symmetry which is originally due to a 
magnetic field. 
Namely this is invariant under translations because it is described by the 
conserved momentum variables. 
Furthermore this is invariant under the translations in the momentum 
variables, $\bf p\rightarrow p+K$, where $\bf K$ is a constant 
vector.\cite{imo} 
The one-particle energy in the lowest order is independent of {\bf p} and 
preserves K-symmetry. 
The density operator is transformed as,
\begin{equation}
\rho({\bf k})\rightarrow e^{i{a\over 2\pi}{\hat k}_x K_y}\rho({\bf k}),
\end{equation}
under $b_l({\bf p})\rightarrow b_l({\bf p}+{\bf K})$
and the interaction Hamiltonian preserves K-symmetry. 
Conversely it is necessary for this symmetry to be broken if one-particle 
energy depends on momentum. 
Perturbative treatment of the interaction Hamiltonian does not give 
momentum dependent one-particle energy. 
The formation of Fermi surface and Hall gas state cannot be studied 
by the perturbation theory. 
We apply, hence, Hartree-Fock mean field theory in this paper. 

\subsection{Compressible mean field solution and symmetry breaking}
Although the mean field solution has been obtained before, 
we present the 
derivation and the solution of compressible Hall gas for completeness. 

We apply a mean field theory and find Hall gas state which has Fermi 
surface in which K-symmetry is broken.  
We study states which preserves 
K-symmetry partly and matches with the Hall bar geometry. 
This is an anisotropic Hall gas which has a momentum dependent one-particle 
energy. 
Self-consistency is required to two point functions and the mean field 
Hamiltonian defined from Eq.~(\ref{hami}),
\begin{eqnarray}
\langle b^\dagger_l({\bf p})b_{l'}({\bf p'})\rangle&=&(2\pi)^2\delta_{ll'}
\sum_N \delta({\bf p}-{\bf p'}-2\pi{\bf N})\\
&&\times\theta(\mu-\varepsilon_l({\bf p}))
e^{-i\phi(p,N)},\nonumber
\end{eqnarray}
where $\mu$ is the chemical potential and $\varepsilon_l$ is the 
one-particle energy given below. 
The phase in Eq.~(\ref{ham}) is a characteristic factor in a 
strong magnetic field and gives important effects generally. 
Now, this phase factor cancels in the translationally invariant mean field 
states but plays important roles in response to external 
potentials and will be discussed later. We find a $K_x$ invariant solution. 
The one-particle energy is given by, 
\begin{equation}
\varepsilon_l({\bf p})=-\int_{\bf BZ}{d^2p'\over(2\pi)^2}
\sum_{\bf N}\tilde v_l({\bf p}'-{\bf p}+2\pi{\bf N})
\theta(\mu-\varepsilon_l({\bf p}')),
\label{sel}
\end{equation}
where $\tilde v_l(\hat{\bf p})=f_{ll}(p/a){\tilde V}(p/a)$. 
Then, the mean field Hamiltonian reads
\begin{eqnarray}
H_{\rm mean}&=&\int_{\rm BZ}{d^2p\over(2\pi)^2}\varepsilon_l({\bf p})
b_l^\dagger({\bf p})b_l({\bf p})\\
&&-{1\over2}\int_{\rm BZ}{d^2p\over(2\pi)^2}
\varepsilon_l({\bf p})\theta(\mu-\varepsilon_l({\bf p})). \nonumber
\end{eqnarray}
The Fermi sea for the $K_x$ invariant Hall gas state is shown in Fig.~1. 

This solution breaks translational invariance in the momentum, 
$p_y$, and is invariant under translation in the momentum, $p_x$. 
The $p_x$ direction is like the integer quantum Hall state which has 
the energy gap of Landau levels but in $p_y$ direction there is no energy gap. 
The slope of the energy at the Fermi energy actually diverges very weakly.

\subsection{Energy per particle, pressure, and compressibility}

The energy per particle, when the energy is measured from the center of each 
Landau levels, of the anisotropic compressible 
state around the filling factor $\nu=n+\nu'$ are given in Fig.~2. 
From these values, the pressure and compressibility of the Hall gas are 
computed and are given in Fig.~3. 
In the present calculation, the charge neutrality is kept and zero momentum 
term of the density is removed in the interaction Hamiltonian 
Eq.~(\ref{hami}). 
The asymmetric parameter $r$ is determined so as to 
minimize the total energy and shown in Fig.~4. 
The pressure and compressibility become negative as are expected. 
Negative pressure and compressibility does not lead to the instability of the 
real system where ions do not move but the electrons move. 
If the positive 
charge as well as negative charge could move, the system may collapse. 
In the real material the positive charge does not move, but only the charge 
carriers move. 
We study its implications later. 
\subsection{Density profile}

To find the density profile of the Hall gas state we express the density 
operator in the present representation. 
That is given as,
\begin{eqnarray}
\rho({\bf x})&=&\int{d^2k\over(2\pi)^2} 
e^{-i{\bf k}\cdot{\bf x}}\sum_{ll'}\int_{\rm BZ}{d^2p\over
(2\pi)^2}b^\dagger_l({\bf p})b_{l'}({\bf p}-a{\hat{\bf k}})
\nonumber\\
&&\times f_{ll'}({\bf k})\exp[-{ia\over4\pi}{\hat k}_x(2p_y-a {\hat k}_y)].
\end{eqnarray}
It should be noted that the phase factor in the above equation depends on 
product between momenta in $x$ direction and $y$ direction. 
The expectation value of the density of Hall gas state which is invariant 
under $K_x$ and has Fermi surface of Fig.~1 becomes as,
\begin{eqnarray}
&\langle \rho({\bf x})\rangle=\rho_0+\sum_{N_x\neq0}
e^{-i{2\pi\over ra}N_x x} {\tilde c}(N_x),\\
&{\tilde c}(N_x)=\int{dp_y\over2\pi}e^{-iN_x (p_y+\pi)}
f_{ll}({2\pi N_x\over r})
\nonumber
\end{eqnarray}
where $p_y$ integration region is made in the filled state region. 
Obviously the density is uniform in $y$ variable but is periodic in $x$ 
variable with a period $ra$. 
Consequently the density profile of the Hall gas is periodic in the 
K invariant 
direction and is uniform in the perpendicular direction.\cite{k} 
This reverse of direction is caused by the phase factor in Eq.~(\ref{ham}). 

\subsection{External potential modulation}

The present system has the rotational invariance. 
Hence the anisotropic Hall gas states have degeneracy with respect to 
the direction of stripe. 
In the real system which has a finite size and disorders, the degeneracy 
is resolved by these effects. 
We study how the degeneracy is resolved by the one-dimensional periodic 
external potential. 
We use the $K_x$ invariant Hall gas state whose density is uniform in 
$y$ direction. 
We fix the direction of modulation of the stripe and 
vary the direction of the modulation of the external potential 
in the following calculation. 
Of course, this is equivalent to fix the external potential and vary 
the direction of modulation of the external potential. 
We see that the energy has a negative correction when both modulations are 
orthogonal but has no correction when both modulations are parallel. 
An external potential modulation of large wavelength, which is 
much larger than the magnetic length, is studied. 
Let us study the perturbative term described by the 
interaction Hamiltonian, 
\begin{eqnarray}
H_{\rm int}&=&\int d^2x \rho({\bf x})v_{\rm ext}({\bf x}),\\
v_{\rm ext}({\bf x})&=&g\cos {\bf Q}\cdot{\bf x},\ 
{\bf Q}=Q(\cos\theta,\sin\theta),\nonumber
\end{eqnarray}
where $v_{\rm ext}({\bf x})$ is the external c-number function and $\rho$ is 
the density operator in a magnetic field. 
It is convenient to express density operator in the momentum space,
\begin{equation}
H_{\rm int}={g\over2}\{\rho({\bf Q})+\rho(-{\bf Q}))\}.
\label{hint}
\end{equation}
The lowest order correction is given as,
\begin{equation}
\Delta E^{(1)}=\langle\Psi_{\rm gas}\vert H_{\rm int}\vert
\Psi_{\rm gas}\rangle
\end{equation}
where $\vert\Psi_{\rm gas}\rangle$ is a many-body state for the 
Hall gas state. 
The Hall gas states are eigenstates of momentum and the wave length of 
$v_{\rm ext}({\bf x})$ is much larger than the lattice spacing $a$. 
Hence the expectation value vanishes. 
The degeneracy is not resolved in the lowest order. 

The second order corrections per area are given as,
\begin{eqnarray}
\Delta E^{(2)}=-\int{d^2p\over(2\pi)}({g\over2})^2&\{&
\vert f_{ll}({\bf Q})\vert^2{1\over\varepsilon_l
({\bf p}+a{\hat{\bf Q}})-\varepsilon_l({\bf p})}
\label{secc}\\
&&+\vert f_{ll}(-{\bf Q})\vert^2{1\over\varepsilon_l
({\bf p}-a{\hat{\bf Q}})-\varepsilon_l({\bf p})}\}.
\nonumber
\end{eqnarray}
The $\bf p$ in the above integral is in the inside of Fermi surface and 
${\bf p}+a\hat{\bf Q}$ or ${\bf p}-a\hat{\bf Q}$ is 
in the outside of the Fermi surface. 
The kinetic energy $\varepsilon_l({\bf p})$ is approximately box-shaped in the 
higher Landau levels.\cite{k} 
The second order correction per area for $K_x$ invariant Hall gas state is 
given approximately as,
\begin{equation}
\Delta E^{(2)}=-{g^2\over4\pi}\vert f_{ll}({\bf Q})\vert^2 
{aQ\over r\Gamma_l}\vert\sin\theta\vert,\label{sec}
\end{equation}
if unscreened Coulomb potential is used. 
Screened Coulomb potential will be discussed in the next subsection. 
Here $\Gamma_l$ is a band width of the $l$ th Landau level. 
$\vert f_{ll}({\bf Q})\vert^2$ is almost unity for a small 
wave number $Q$. 
The second order correction vanishes at $\theta=0$. 
The product of any operators that has momentum in $p_x$ direction 
always vanishes for $K_x$ invariant Hall gas state. 
The degeneracy is resolved and the energy becomes minimum at 
$\theta =\pi/2$. 
Hence the $K_x$ Hall state has lower energy and is realized in the system 
with $y$ dependent density modulation. 
$K_x$ invariant Hall state has uniform density profile in $y$ direction 
and periodic density profile in $x$-direction. 
Hence the Hall gas state which has perpendicular density profile to external 
potential modulation is realized as shown in Fig.~5. 
This orthogonality is caused by the phase factor in the density 
operator\cite{pb} and agrees with the recent experiment.\cite{Willett} 

Due to the effect of the residual interaction $H_{\rm res}$, 
which is obtained from the normal ordered product of $H_1$ in 
Eq.~(\ref{hami}) with respect to the Hartree-Fock ground state, 
the energy correction is modified slightly. 
When we apply the same approximation on the shape of one-particle 
energy spectrum, 
we see that the following term are added into the energy 
\begin{equation}
-{11\over1024r^2\Gamma_l}({q^2\over a})^2+{g^2 a 
Q\over4\pi r^2 \Gamma_l^2}
\vert f_{ll}({\bf Q})\vert^4\vert \sin\theta\vert {q^2\over a}. 
\end{equation}
The first term is independent of the external modulation 
and simply shifts the ground state energy. 
But the second term, 
which is the first order term in $H_{\rm res}$, has a same dependence on 
external modulation as the lowest order term and is smaller than the first 
order term by factor five for $\nu=2+1/2$. 
From the next order term, which is second order in $H_{\rm res}$, 
the energy has a different dependence on the external modulation and 
does not vanish even if the angle $\theta=0$. 
An angle independent term is generated when 
$H_{\rm res}$ is multiplied first to the 
Hartree-Fock ground state and the Hamiltonian of external modulation 
Eq.~(\ref{hint}) is multiplied next. 
This term does not vanish but the magnitude is very small compared to the 
zeroth term Eq.~(\ref{sec}). 
Thus higher order corrections from the residual interaction are small, 
although the energy correction at $\theta=0$ does not vanish. 
Consequently, orthogonal phase has a lower energy than the parallel phase 
even though higher order corrections from the residual interaction are 
taken into account.
Our results are consistent with Yoshioka's results\cite{Yoshi} 
based on the numerical calculation for the charge density wave 
qualitatively but give a deep insight for the reason of the 
phenomena.\cite{Ao}

In the experiments,\cite{e,f} it is suggested that the direction of the 
stripe order tends to be perpendicular to in-plane constant magnetic 
fields. 
This is consistent with the Hartree-Fock calculation.\cite{i,j} 
Using the results obtained above, we can investigate the case of 
in-plane periodic magnetic fields. 
For example, in addition to the perpendicular magnetic field 
${\bf B}=(0,0,B_z)$, 
we consider the in-plane periodic magnetic field 
${\bf B}_{\rm in}=(B_{\rm in}\cos Q y,0,0)$. 
The vector potential for this in-plane magnetic field is given by 
${\bf A}_{\rm in}=(0,0,B_{\rm in}(\sin Q y)/Q)$. 
If we treat the additional term for ${\bf A}_{\rm in}$ as a perturbation, 
the same conclusion is obtained as the case of the periodic potentials.
The second order correction becomes minimum when the stripe is 
uniform in the $y$ direction, which is perpendicular to the in-plane magnetic 
field. 
This result is consistent with the experiment of in-plane constant magnetic 
fields in $Q\rightarrow 0$ limit. 

\subsection{Coulomb screening}

In the quantum Hall system the kinetic energy is frozen and 
Coulomb potential is not screened in the lowest order. 
Since the kinetic energy is generated by the unscreened Coulomb potential, 
Coulomb potential might be screened. 
If it happens, anisotropic Hall gas may have different properties. 
It is important to know if the Coulomb potential is screened. 
We study an implication of the screening and if the screening occurs in 
the self-consistent manner.

If the Coulomb potential is screened, Fermi velocity becomes finite, and 
one-particle energy is given as,
\begin{eqnarray}
\varepsilon({\bf p}_F+{\bf q})=\varepsilon({\bf p}_F)+v_F q_y,\\
v_F={\partial\varepsilon\over\partial p_y}\Bigr\vert_{
{\bf p}={\bf p}_F}. 
\nonumber
\end{eqnarray}
The denominator in the second order energy correction (\ref{secc}), 
is proportional to $q$ 
and the integration region is also proportional to $q$. 
Hence the second order correction due to long wavelength 
potential modulation becomes,
\begin{equation}
\Delta E^{(2)}=-({g\over2})^2\vert f_{ll}({\bf Q})\vert^2 
{1\over\pi\vert v_F\vert},
\end{equation}
which is very different from Eq.~(\ref{sec}) and has no angle dependence. 
Thus the energy correction is independent of the orientation of 
compressible states and the degeneracy is not resolved, when the Coulomb 
potential is screened. 
Since the degeneracy stays, the periodic external density modulation 
does not stabilize anisotropic Hall state of a particular orientation.
Perhaps a symmetric state which is a linear combination of states of
different orientations could be realized in this case. 
Then anisotropic state would not be observed. 

Now, we study if a self-consistent screened solution exists. 
To find a self-consistent screened solution, 
we start from a screened Coulomb potential $\tilde v_{\rm TF}$ 
\begin{equation}
\tilde v_{\rm TF}({\bf p})={1\over \tilde v_l({\bf p})^{-1}+m_{\rm TF}}.
\end{equation}
We substitute this potential into the self-consistency 
condition (\ref{sel}) 
and obtain self-consistent kinetic term for the electrons first. 
Here $m_{\rm TF}$ is the Thomas-Fermi (TF) screening mass. 
From the one-particle energy thus obtained, 
new TF screening mass is calculated by the Fermi velocity $v_F$ as 
$m_{\rm TF}=1/\pi a v_F$. 
The final value is compared with the original screening mass. 
In Fig.~6, the computed TF screening mass is given as a 
function of the initial TF screening mass for $r=2$. From 
this result, two values always agree at the origin, where screening 
effect vanishes. 
There is a second solution with finite 
screening length in the states of the lowest Landau level. 
In this solution that has a finite screening length, one-particle state at 
the Fermi energy has a finite velocity. 
Hence, in the lowest Landau level, 
the self-consistent screening is possible and 
anisotropic state may not be realized. 

In higher Landau levels, the first solution of zero screening length exists
but the second solution of finite screening length does not exist. 
In the first solution Coulomb force is not screened and the previous 
self-consistent solution is the solution. 
In this solution the orientational
degeneracy is resolved by the periodic external density modulation, and 
the Hartree-Fock state with anisotropic Fermi surface of a particular 
orientation is stabilized. 
Hence in the system of small density modulation,
anisotropic Hall gas is stabilized in high Landau levels but anisotropic Hall 
gas is not stabilized  in the lowest Landau level. 
Instead, the screenning length is finite in the lowest Landau level.   
Our results are consistent with the facts that anisotropic states have been 
observed experimentally only in higher Landau levels and the compressible 
state at the half-filling of the lowest Landau levels is isotropic.

It should be noted that the new TF screening mass becomes negative for the 
half-filled second Landau level in Fig.~6. 
This indicates the instability of the Hall gas state which is consistent 
with the experiment at $\nu=5/2$ where the fractional quantum Hall effect 
is observed.\cite{pan} 
Theoretically it is pointed out that the Hall gas state is unstable to the 
Cooper pairing\cite{pair,npb} at the half-filled second Landau level.

\section{
Shape of the anisotropic Hall gas with a negative pressure}

Gas of negative pressure shrinks and strip of compressible gas is formed. 
A similar strip is formed also in the system of finite sizes and 
disorders due to their interplay.

\subsection{Static Coulomb energy}

Compressible states of negative pressure are deformed. 
Uniform Hall gas of the 
previous section is composed of electrons and uniform neutralizing 
background charges, which are composed of ions. 
The electrons are movable but the ions are not movable. 
Consequently, when electrons shrink, charge neutrality is violated 
partly and static Coulomb energy is added to the system's energy. 
We study this static Coulomb energy here.   

We study how much the compressible gas of the filling factor $n+\nu'$ shrinks
in the system of length $L$ and the width $L_w$ due to negative pressure. 
We assume that the gas shrinks uniformly by $\delta_x$ in the width.

The static Coulomb energy per particle of this situation is given by,
\begin{equation}
E_{\rm Coulomb}=\nu' {L_w\over a} g({\delta_x\over L_w}){q^2\over a},
\end{equation}
where $g(x)$ is a monotonically increasing function which 
behaves as $-{x^2\over2}\log x$ in $x\rightarrow 0$ and diverges at $x=1$ 
logarithmically. 
This energy is added to the energy of the neutral system and the total energy 
of the system is obtained. 
The total energy is expressed as a function of the $\delta_x$ in Fig.~7. 
The energy becomes minimum at one value of the $\delta_x$.

At the above value the Hall gas is stabilized and the pressure vanishes
and the compressibility becomes positive. 
The value depends on the filling 
factor. 
$\delta_x$ nearly vanishes when the filling of the Hall gas is 
large, and $\delta_x$ becomes finite if the filling is small. 
Hence the Hall gas is confined to strip state if the filling is slightly 
away from integer. 

\subsection{Interplay between localization length and finite size}

As was pointed out in the introduction, long range disorder potentials 
which are realized in GaAs systems are studied in the present paper. 
Disorder effects in finite system are analyzed in this subsection. 
Localized 
states due to disorders have discrete energies and finite localization lengths.
These states have normalizable wave functions that vanish at infinite spatial 
region. 
Hence electrons in these states neither carry the bulk electric current 
nor contribute to electric conductances in the infinite system. 
In finite systems, however, localized 
states could behave like extended states if localization lengths are
larger than or equal to the system size. 
They contribute to the electronic 
conductances if they have finite wave functions at current contact regions. 
Thus it is convenient to classify the one-particle energy 
region to several regime depending upon their localization lengths and the 
systems' sizes. 
There are three systems' sizes. 
One is the length of the system, $L_s$, the second one is the width in the 
Hall probe region, $L_{w1}$, and the third one is the width in the potential 
probe region, $L_{w2}$. 
Generally the first one is the largest, the second one is medium, and the 
third one is smallest.

The localization length depends on the state's energy and we write it as 
$\lambda(E)$. 
$\lambda(E)$ diverges at the center of the Landau levels as\cite{aoki} 
\begin{equation}
\lambda(E)=\lambda_0\vert E-E_l\vert^{-s},\ 
s\approx2,
\end{equation}
and becomes minimum at the middle between two Landau levels as is given in 
Fig.~8. 
Three energy values correspond to three lengths. 
Let $E_s$, $E_{w1}$, and $E_{w2}$ be the energy values where the 
localization lengths agree with the three system's sizes, 
$\lambda(E_s)=L_s$, 
$\lambda(E_{w1})=L_{w1}$, 
$\lambda(E_{w2})=L_{w2}$. Then states in the energy range $\vert E-E_l
\vert<E_s$ are extended states, 
and we call this energy region as the extended state region. 
States in the energy range, $E_s<\vert E-E_l\vert<E_{w1}$, 
bridge from one edge to the 
other edge at the potential probe region and at the Hall probe region, 
and we call this energy region as the collapse regime. 
States in the energy range, $E_{w1}<\vert E-E_l\vert<E_{w2}$, 
bridge from one edge to the other edge at the potential probe region, 
and we call this energy region the dissipative quantum Hall regime (QHR). 
Finally states in the energy range, $E_{w2}<\vert E-E_l\vert$, are localized, 
and we call this energy region as the localized states region or QHR. 
Reasons why we use particular names for the last two regions will 
become clear later. 
The energy regions and typical wave functions in configuration space are 
written in Fig.~9.

\subsection{Strip formation}

There are two mechanisms of forming strip of electrons. 
First one is due to interactions. 
One-particle wave functions are extended if its energy is in the center of 
Landau levels and cover whole spatial region from one end 
to another end in the system of no interactions. 
Interaction, however, modifies the many-body wave functions and compressible 
Hall gas of negative pressure is formed. 
Due to negative pressure, which is 
obtained in the previous section, this Hall gas shrinks. 
The strip of compressible states is formed. 
They have energies near the center of Landau levels. 

Second one is due to disorders in the finite systems. 
Wave functions in the collapse regime and dissipative QHR 
bridge from one edge to another edge and are like stripe even 
without interactions. 
Thus the strip is formed in the extended states region, 
in the collapse regime, and in the dissipative QHR. 
The electric resistances of the strip state are studied in the Sec.~V.    

\section{Transport properties}

In this section, the transport of the system when the anisotropic 
Hall gas fills whole spatial regions is studied. 

\subsection{Current basis and topological formula of Hall conductance}

It is convenient to define a new basis called current basis,\cite{l,m} 
where electron operators satisfy naive Ward-Takahashi identity between the 
vertex part and the propagator, first. 

Let us define a unitary matrix 
and transform the electron operators $b_l$ and the propagator $S_{ll'}$: 
\begin{eqnarray}
U^\dagger_{ll'}({\bf p})&=&\langle f_l\vert e^{i({\tilde p}_x\xi+
{\tilde p}_y\eta)/a-{i\over4\pi}p_xp_y}\vert f_{l'}\rangle,\label{uni}\\
{\tilde b}_l({\bf p})&=&\sum_{l'}U_{ll'}({\bf p})b_{l'}({\bf p}),
\nonumber\\
{\tilde S}(p)&=&U({\bf p})S(p)U^\dagger({\bf p}).
\nonumber
\end{eqnarray}
Electric current are expressed with these operators as
\begin{eqnarray}
j^\mu({\bf k})&=&\int_{\rm BZ}{d^2p\over(2\pi)^2}\sum_{ll'}
{\tilde b}_l^\dagger({\bf p})v^\mu{\tilde b}_{l'}({\bf p}-a\hat{\bf k}),
\nonumber\\
v^\mu&=&(1,-\omega_c\eta,\omega_c\xi)\\
&&+{a\omega_c\over2\pi}(0,\tilde p_x-{ak_x\over2},
\tilde p_y-{ak_y\over2}),
\nonumber
\end{eqnarray}
where $\tilde{\bf p}=(p_x/r,rp_y)$. 
The charge density satisfies the commutation relation:
\begin{equation}
[\rho({\bf k}),{\tilde b}_l({\bf p})]\delta(t-t')=
-{\tilde b}_l({\bf p}-a\hat{\bf k})\delta(t-t').
\label{new}
\end{equation}
Thus the electron operators satisfy the commutation relation of 
local field theory in each Landau level. 
This gives important and 
powerful Ward-Takahashi identity and other relations. 
From this relation the following Ward-Takahashi identity between the vertex 
part and the propagator of the matrix form is derived,
\begin{eqnarray}
&{\tilde\Gamma}^\mu(p,p)=t^{\mu\nu}{\partial{\tilde S}^{-1}(p)\over\partial 
p^\nu},\label{wt}\\
&t^{\mu\nu}={\rm diag}(1,ar,a/r). 
\nonumber
\end{eqnarray}
The Hall conductance is expressed by the topological invariant 
expression,\cite{l,m}
\begin{eqnarray}
\sigma_{xy}={e^2\over h}{1\over24\pi^2}\int_{\rm BZ\times S^1}&d^3p
&\epsilon^{\mu\nu\rho}{\rm tr}(\partial_\mu{\tilde S}^{-1}(p){\tilde S}(p)
\label{topo}\\
&&\times\partial_\nu{\tilde S}^{-1}(p){\tilde S}(p)
\partial_\rho{\tilde S}^{-1}(p){\tilde S}(p)).
\nonumber
\end{eqnarray}
In the QHR, this formula gives quantized value $n{e^2\over h}$. 
In the finite systems the momentum becomes discrete, and the expression 
(\ref{topo}), is valid and the value is quantized exactly as well.\cite{br} 
The systems of many-body interactions is treated by perturbative expansion.
Since the current conservation and the commutation relations between 
the charge density and the field operators are kept intact in the system 
of interaction, the full propagator and full vertex part which include 
interaction effects satisfy Ward-Takahashi identity, Eq.(\ref{wt}). 
Hence the renormalized quantities also satisfy the same Ward-Takahashi 
identity and the topological formula of the Hall conductance,\cite{l,m} 
is written by the renormalized propagator in the current basis. 
In the QHR there is no extended states at Fermi energy and 
the Hall conductance stays at the exact quantized values. 
The value is not modified by the interactions. 

Now we study the Hall gas regime. 
The free Hamiltonian, $H_0$ is given by the non-diagonal form,
\begin{equation}
H_0=\sum{\tilde b}_l^\dagger({\bf p})\{U_{ll'}({\bf p})E_{l'}
U^\dagger_{l'l''}({\bf p})\}{\tilde b}_{l''}({\bf p}). 
\end{equation}
Using the energy eigenvalues of the previous section for the Hall gas we
find the propagator in the current basis. 
The propagator is given around the Fermi energy as 
\begin{equation}
S^{(c)}_{ll'}(p)={\delta_{ll'}\over p_0-(E_l+\varepsilon_l({\bf p}))}
\label{prop}\\
\end{equation}
in the energy basis, where $\varepsilon_l({\bf p})$
was computed in Section 2.
We substitute  Eq.(\ref{uni}) and Eq.(\ref{prop}) to 
Eq.(\ref{topo}), and we have 
\begin{equation}
\sigma_{xy}={e^2\over h}(n+\delta n),
\end{equation}
at $\nu=n+\delta n$. 
Thus the Hall conductance is proportional to the filling factor 
in the Hall gas regime. 

\subsection{Longitudinal resistances}

The $K_x$-invariant anisotropic Hall gas states at $\nu=n+1/2$ have the Fermi 
surface parallel to $p_x$ axis. 
In $p_x$ direction, there is no empty state within the same Landau level. 
This direction is like the integer quantum Hall state. 
Thus, longitudinal conductance in this direction vanishes. 
In the $p_y$ direction, the system behaves like one-dimensional.\cite{n} 
We assume that the small voltage probe contact is attached in the edge region. 
The current in this region is due to one-dimensional edge mode which has 
$p_y$-dependent one-particle energy in addition to Hall electric field 
dependent one-particle energy. 
The latter energy corresponds to one chiral mode 
which does not dissipate energy and gives the ordinary Hall resistance 
as were shown by Halperin\cite{Hal} and Buttiker\cite{Bu} 
but the former one-particle energy 
corresponds to both chiral modes and dissipate energy. 
We apply Landauer formula to this part of the energy. 
The velocy is given as,
\begin{equation}
v_y={\partial \varepsilon(p_y)\over\partial p_y},
\label{v}
\end{equation}
and number of states of the extended states in the chemical potential 
difference $\Delta\epsilon$,
\begin{equation}
\Delta n={1\over2\pi}{\partial p_y\over\partial\varepsilon}
\Delta\varepsilon (1+C),
\label{C}
\end{equation}
where $C=(\partial\delta_p/\partial p)_{p_F}/L$ is the correction term 
which comes from the phase shift $\delta_p$ due to impurity scattering. 
Combining Eqs.~(\ref{v}) and (\ref{C}), we have total current
\begin{eqnarray}
I_y&=&ev_y\Delta n\nonumber\\
&=&{e^2\over2\pi}(1+C)V_y,
\end{eqnarray}
where $V_y$ is voltage in $y$ direction and  the chemical potential difference
$\Delta\epsilon$ is replaced with $eV_y$. 
The precise value of the correction term $C$ is not known. 
It is, however, expected that $C\rightarrow-1$ when the Landau level 
index goes to infinity from the scaling theory of Anderson localization 
in the system of zero magnetic field, and $C$ vanishes in the 
ideal system with no impurities. 
We have, hence, conductance
\begin{equation}
\sigma_{yy}={e^2\over h}. 
\end{equation}
The resistance in $x$ direction becomes finite and vanishes in $y$ 
direction. 
If the compressible states carry a small amount of current, which occurs 
when the filling factor slightly deviates from integer, the above value 
gives the proportional constant between the voltage and the current 
generated by the compressible states. 

\section{Current activation into a strip of compressible states from 
undercurrent}

The transport properties of the strip of compressible Hall gas
which covers a small part of the whole system and is disconnected from 
current contacts are studied in this section. 
If the current is induced in this strip of 
compressible states, this current generates longitudinal resistance. 
Whereas 
in other part which has no compressible states, Hall resistance is quantized. 
Thus a new quantum Hall regime where the Hall resistance is quantized exactly 
in the system of finite longitudinal resistance is derived. 

\subsection{Current activation from undercurrent as a new tunneling mechanism}

As shown in Sec.~III, strip formed from localized electrons of localization 
lengths between the width and length  of the system in which the width is 
smaller than length does not reach at least to one 
current contact 
and injected current does not flow through  the strip at zero temperature.
Strip in this situation has peculiar 
transport properties and gives important effects to the 
IQHE with finite injected currents at finite temperature.

We study the situation where several lower Landau levels are filled completely 
and one Landau level is partially filled as shown in Fig.~10. 
The current-carrying strip is formed in the highest Landau level and 
bridges one side to the other side. 
Although Fermi energy is in the strip state energy region, 
the strip is disconnected from the current contacts. 
Injected electric current flows through the lower Landau levels, 
and does not flow through the isolated compressible states of higher Landau 
levels if there is no interaction. 
In reality the interaction exists and modifies the isolated states. 
At finite temperature the current is activated into the isolated strip 
states from lower Landau levels as shown in fig.~10. 
We study the effects of Coulomb interaction at finite temperature
and find that the strip states have small amount of 
electric current at low temperature. 
The magnitude of induced current depends on injected current as well 
as on temperature. 

From Fermi-Dirac statistics, particles are filled first in the 
lowest energy level 
and second in the next levels and so on. 
The dissipation occurs only near the Fermi surface and does not occur at 
the lower energy levels. 
In the quantum Hall system, the bulk current is carried by the electrons 
in lower levels and in the levels near Fermi surface. 
So in the present situation shown in Fig.10, dissipationless 
current flows in the lower levels and induced current in the stripe in the 
middle of sample near Fermi surface dissipates energy. 
This is very different from and should not be confused with ordinary 
parallel circuit.

The second order correction from the Coulomb interaction which is expressed 
in Feynman diagram of Fig.~11 gives the induced current to the strip states. 
One electron line (double line) stands for those of lower Landau levels 
which are connected with current contacts and are carrying current. 
Thus these states depend on the Hall electric field.\cite{br} 
The other line (solid line) stands for those of strip states around Fermi 
energy which are disconnected from current contacts and are not carrying 
current in the lowest order. 
The dashed line stands for the Coulomb interaction. 
To compute induced electric current, the current operator is 
inserted in the strip state line and is shown as wavy line in Fig.~11. 
We express operators, propagators, vertices, and others in 
the current basis. 
The induced current is calculated as
\begin{eqnarray}
j_{\rm ind}&=&{1\over\beta}\sum_{\omega=i\omega_n}
\int{d^2 p d^2k\over(2\pi)^4}{\rm Tr}(
{\tilde V}({\bf k}){\tilde S}^{(c)}(p-ak)\\
&&\times{\tilde \Gamma}^x(p-ak,p-ak){\tilde S}^{(c)}(p-ak){\tilde S}(p;
E_H))
\nonumber\\
&=&{1\over\beta}\sum_{\omega=i\omega_n}
\int{d^2 p d^2k\over(2\pi)^4}{\tilde V}({\bf k})
\vert f_{l,l+1}({\bf k})\vert^2 \nonumber\\
&&\times\partial_x S_{l,l}(p;E_H)S^{(c)}_{l+1,l+1}(p-ak)\nonumber
\end{eqnarray}
where $S(p;E_H)$ is the propagator which depends on the Hall electric field 
$E_H$ and $S^{(c)}(p)$ is the propagator for the stripe states which does not 
depend on $E_H$. 
$\omega_n=(2n+1)\pi/\beta$ is the Matsubara frequencies. 
The chemical potential $\mu$ is in between $E_l$ and $E_{l+1}$. 
In the above calculation, we considered only the mixing term between 
the Landau levels $E_l$ and $E_{l+1}$ which is the dominant term in the 
induced current. 
The asymmetric parameter is fixed at $r=1$ for simplicity. 
Then the induced current reads
\begin{eqnarray}
j_{\rm ind}&=&\int{d^2 p d^2k\over(2\pi)^4}{\tilde V}({\bf k})
\vert f_{l,l+1}({\bf k})\vert^2 v_x\beta \\
&&\qquad\qquad\times e^{\beta\Delta E_1({\bf p}-a{\bf k})}\theta(-\Delta 
E_2({\bf p})),\nonumber\\
\Delta E_1({\bf p})&=&E_l+eE_H^{\rm eff}{a\over2\pi}p_x-\mu,
\label{eh}\\
\Delta E_2({\bf p})&=&E_{l+1}+\varepsilon_{l+1}({\bf p})-\mu,\nonumber\\
v_x&=&eE_H^{\rm eff}{a\over2\pi},\nonumber
\end{eqnarray}
where $E_H^{\rm eff}$ is an effective Hall electric field $E_H^{\rm eff}=
E_H/\gamma$, which is enhanced by the strong localization effect.\cite{br} 
In experiments,\cite{kawa} $1/\gamma$ is on the order of 10. 

The induced current is given by,
\begin{eqnarray}
&j_{\rm ind}=V_l e^{\beta(E_l-\mu)}\sinh{\beta e E_H^{\rm eff}\over 2},
\label{act}\\
&V_l={4\pi\nu'q^2\pi^{3/2}\over a^2 (l+1)}\int_0^\infty dx x^{1/2}\{L_l
^{(1)}(x)\}^2 e^{-x}.
\nonumber
\end{eqnarray} 
The value depends on the Hall electric field, the total current, and the 
temperature. 
The temperature dependence is of activation type and the gap energy is 
linear in $E_H$.

\subsection{Induced current in a strip of compressible Hall gas}

The temperature dependence of the current is of activation type. 
At relatively high temperature, the electron temperature is the same as 
that of the whole system. 
At low temperature, however, the electron temperature is different from the 
system's temperature and is determined from the energy due to 
the electric resistance and the heat conduction from the electron system to 
the lattice system. 
The temperature determines the electric resistance and conversely the 
electric resistance determines the temperature. 
Hence they are determined self-consistently.\cite{komiyama} 

The induced current of the strip depends on the total injected current 
through the Hall electric field. 
Effective band width (\ref{eh}) of extended Landau levels 
in the system of Hall electric field increases with Hall electric field. 
This agrees with the experiments of the breakdown of the IQHE due 
to the injected current, where the activation energy 
depends on the injected current. 
So we use these previous results in the present work. 

\subsection{Longitudinal resistance}

The total longitudinal resistivity is the ratio between 
the induced electric field in the compressible gas region and the 
current density and is given as,
\begin{equation}
\rho_{xx}=\rho^{(s)}_{xx}ej_{\rm ind}/\sigma_{xy}E_H,
\label{rho}
\end{equation}
where $\rho^{(s)}_{xx}$ is the longitudinal resistivity of the stripe state. 
The particular dependence of the longitudinal resistance on the Hall field 
is obtained for the first time. 
This result are compared with the experiment of Kawaji et al.\cite{b} 
in Fig.~12. 
In the experiment, the filling factor is 4, magnetic field is 5.7 T, and 
the temperature is 0.75 K. 
The electron temperature could be different from the above value due to 
heating effect and is not known experimentally. 
So we fit the theoretical curve by changing the temperature, $\gamma$, 
and $\nu'$. 
As shown in Fig.~12, reasonable agreements are obtained when the 
temperature is 1.576 K, $\gamma=0.065$, $\nu'=0.5$ 
and $\mu=E_l+\hbar\omega_c/2$. 
This temperature is that of the electron system at the strip and 
is different from the experimental value which is at the lattice system. 
The large deviation at large Hall electric fields is caused by 
the breakdown of IQHE. 

\subsection{Dissipative QHR}

The transport properties of the finite system depends on the system sizes, 
the position of the Fermi energy, and the magnitude of the induced current. 
If the width in the Hall probe region is wider than the width in the 
potential probe region, longitudinal resistance and Hall resistance 
behave very differently. 
We see that breakdown should occur in two steps.
 
As the current increases, mobility edges move and the localization length 
increases. 
When the localization length of a state reaches the width of only the 
potential probe region, this state contributes to longitudinal resistance but 
not to the Hall resistance. 
Potential probe region has an temperature dependent electric 
resistance of activation type. 
The Hall probe region has still only localized states which have shorter 
localization lengths than the width. 
Hence the Hall probe region is considered as the QHR in this region. 

The Hall resistance measured in the wider Hall probe region becomes 
that of the finite 
system in the localized state region where momentum variables become discrete.
As was shown in  Eq.~(\ref{topo}), the Hall 
resistance is written as a peculiar topological invariant form in the 
momentum space. 
The integrand is determined by the magnetic field and has no 
dependence on spatial component of the momentum. 
Hence this topological 
invariant does no change the value even when the integration variables are
replaced with those of finite system, discrete values. 
Thus this has no finite 
size corrections in the localized state region.\cite{br} 
Hence the Hall resistance measured in the present situation 
agrees with the exactly quantized value.
Thus resistivities at low temperature are given in our theory as,
\begin{eqnarray}
\rho_{xx}&=&\tilde\rho e^{-\beta\Delta E_{\rm gap}},\nonumber\\
\rho_{xy}&=&({e^2\over h}N)^{-1},\label{diss}\\
\Delta E_{\rm gap}&=&\mu-E_l-eE_H^{\rm eff}a/2,\nonumber
\end{eqnarray}
where $\tilde\rho$ is temperature independent.
$\rho_{xy}$ is quantized even though $\rho_{xx}$ does not vanish. 

Eq.~(\ref{diss}), show that the Hall resistance is quantized 
even though the longitudinal resistivity does not vanish. 
This is a new regime of IQHE, which has not been 
expected from the naive picture of IQHE. 
The Eq.~(\ref{diss}) shows that the new dissipative QHR is realized only in 
the finite system.  

If the width in the Hall probe region is same as the width in the 
potential probe region, the dissipative QHR does not exist. 

\subsection{Collapse and breakdown of IQHE}

If the current increases further the extended state regions are broadened 
in energy by the amount that is proportional to Hall electric field. 
Mobility edges move with Hall electric field towards outward of Landau 
levels. 
The localization lengths of the localized states become larger. 
At some current value, they reach the spatial width of the Hall probe region. 
Wave functions in this energy region bridge both of the potential probe 
region and the Hall probe region. 
The electric current in this region is given by the previous form of the 
activation type in Eq.~(\ref{act}). 
If the Fermi energy is in this region, not only Hall resistance has a small 
correction but also longitudinal resistances becomes finite 
at low temperature. 
They are given in our theory as,
\begin{eqnarray}
\rho_{xx}&=&\tilde\rho e^{-\beta\Delta E_{\rm gap}},\\
\Delta\rho_{xy}&=&{\tilde \rho}'e^{-\beta\Delta E_{\rm gap}}.
\end{eqnarray}
This corresponds to the collapse of IQHE. 
When the current increases further and the localization lengths of the 
localized states become even larger and reach the spatial 
length of the system. 
Wave functions in this energy region cover whole area of the system. 
If the Fermi energy is in this region, 
the Hall resistance and the longitudinal 
resistivity have finite corrections and are given by,
\begin{eqnarray}
\rho_{xx}&=&{h\over e^2}\delta,\\
\Delta\rho_{xy}&=&{h\over e^2}\delta',
\end{eqnarray}
where $\delta$ and $\delta'$ are finite numbers which do not vanish 
in $\beta\rightarrow\infty$. 
This corresponds to the breakdown of the IQHE in the finite system. 
If the current increases further and exceeds a critical value in which 
the energy gap $\Delta E_{gap}$ vanishes completely, 
the localization length becomes infinite, the longitudinal resistance 
becomes finite, and Hall resistance deviates from a quantized value 
substantially. 
The IQHE in the infinite system disappears then. 
The critical value of Hall electric field, $E_c$, is given from 
Eq.~(\ref{eh}) as,
\begin{equation}
{\hbar\omega_c\over2}=eE_c^{\rm eff} a. 
\end{equation}
The critical Hall electric field is proportional to,
\begin{equation}
E_c={\hbar\omega_c\over2e}{\gamma\over a}.
\end{equation}
$E_c$ is proportional to $B^{3/2}$ which is consistent with the 
experiment.\cite{kawa}

\section{Summary}

We have shown that the anisotropic Hall gas obtained from Hartree-Fock mean 
field theory has unusual thermodynamic properties and transport properties.
They agree with the recent experiments of the orientational symmetry breaking 
states at higher Landau levels and collapse phenomena. 
Existence of 
new quantum Hall regime with a finite dissipation, where Hall resistance is 
quantized exactly even with a finite longitudinal resistance, is derived.
These results are connected with characteristic properties of the quantum 
Hall systems. 

\subsection{Non-commutative guiding center variables}

One of the characteristic features of the electron field in quantum Hall 
system is the non-commutative coordinates of guiding center in 
Eq.~(\ref{comm}). 
This property gives following results: 

(1) The commutation relation between the charge density 
(\ref{ham}) and the electron field (\ref{ele}) in the Landau level 
is given by,
\begin{eqnarray}
&[\rho({\bf k}),b_l({\bf p})]\delta(t-t')=-\sum_{l'}f_{ll'}({\bf k})b_{l'}
({\bf p}-a\hat{\bf k})\\
&\qquad\qquad\qquad\qquad
\times e^{-i{a\over4\pi}\hat k_x(2p_y-a\hat k_y)}\delta(t-t').
\nonumber
\end{eqnarray}
As was emphasized in Ref.~(\cite{l}) the expansion of the right hand side 
in the momentum $q$ has a linear term, i.e., the dipole moment. 
Hence Ward-Takahashi identity between the vertex part and the propagator is 
modified from the standard form of ordinary local field theory. 
This term is in fact absorbed in the unitary matrix of Eq.~(\ref{uni}). 
The new field is defined from the old field by the unitary transformation. 
The commutation relation between the charge density and the electron 
field in the current basis is given then by Eq.~(\ref{new}). 
This leads to the standard form of Ward-Takahashi identity Eq.~(\ref{wt}) 
between the vertex part and the propagator. 
The Hall conductance is written by the topologically 
invariant form, Eq.~(\ref{topo}). The value of the topological invariant 
becomes finite and increasing with filling factor due to the effect of the 
transformation matrix. Thus due to the dipole moment, the value of the 
winding number of propagator, the Hall resistance becomes finite.  

(2) The density profile of the $K_x$-invariant anisotropic Hall gas, 
in which one-particle states are completely filled in $p_x$ direction as shown 
in Fig.~1, is uniform in $y$ direction but is periodic in $x$ direction. 
This directional orthogonality between the uniform density in the momentum 
space and that in the coordinate space are due to the phase factor 
in Eq.~(\ref{ham}). 

(3) The ground state is affected and its energy is lowered when the external 
periodic potential modulation which has a wave vector perpendicular to the 
Fermi surface is added as seen in Eq.~(\ref{sec}). 
Thus anisotropic state that has the density profile perpendicular to the 
external potential modulation is stabilized and realized. 
This orthogonality is caused by the phase factor in Eq.~(\ref{ham}). 

(4) Due to the phase factor in the charge density of Eq.~(\ref{ham}) 
static potential which depends upon local coordinate gives bi-local
coordinate dependence to the electron pairs.Thus  electrons move  in 
space and have the energy broadened. 
As was shown in Ref.~(\cite{br} and is given in Eq.~(\ref{eh}), the 
one-particle states in uniform electric field have the band energies 
whose width is determined by the electric field as 
$E_l+eE_H^{\rm eff}{a\over 2\pi}p_x$.
This causes the breakdown due to injected current \cite{br}.

(5) The phase factor in the charge density of Eq.~(\ref{ham}) is proportional 
to the product between 
the total momentum, $k_{\rm total}=2p/a+k$, and relative momentum, $k$. 
Hence the Fourier transform in the relative coordinates, $r$, is a gaussian 
function of $r+a k_{\rm total}$ and decreases fast. 
Hence the relative coordinates couple with the total momentum. 
The relative distance becomes large only for  the large total momentum. 
Because the ionization energy of the quantum Hall system is 
determined from particle-hole pair of large distance, that is given from 
the energy of the particle-hole pair at the large momentum.     

\subsection{Frozen kinetic energy}

Another feature of the quantum Hall system is that the one-particle state
has a discrete energy and the kinetic energy is frozen in the absence of 
interaction. 
This causes the K-symmetry, i.e., the translational invariance in 
the momentum space. 
In the K-invariant state, the Fermi surface does not exist. 
Conversely in the state which has the Fermi surface the K-symmetry is 
broken. 
We have studied the Hall gas state in which the K-symmetry is 
spontaneously broken by the Coulomb interaction and found that the Hall gas 
has anomalous thermodynamic properties such as negative pressure and 
compressibility. 
They cause the strip formation. 
The strip in the bulk is disconnected with the current contacts and the 
transport in the strip occurs through the current activation from 
undercurrent in the lower Landau level. 
Thus the anomalous charge transport occurs at finite temperature. 

\subsection{Dissipative QHR and metrology}

The localized states of having large localization length give equivalent 
effects as the compressible strip. 
Localized states bridge at the potential probe region but not at the 
Hall probe region if their localization length are the same order as or 
larger than the system width in the potential probe region but are smaller 
than those in the Hall probe region. 
They contribute to the longitudinal resistance but do not contribute to the 
Hall resistance. 
Then the Hall resistance is quantized exactly but the longitudinal 
resistance does not vanish. 
This is a new regime which we call the dissipative QHR. 
The existence of the dissipative QHR is important 
in metrology of the IQHE.\cite{metro} 

\acknowledgements

We would like to thank Y. Hosotani, R. Jackiw, W. Pan, and P. B. Wiegmann 
for useful discussions. 
One of the present authors (K. I.) thanks S. Kawaji and B. I. Shklovskii 
for fruitful discussions. 
One of the present authors (N. M.) is grateful for stimulative conversations 
with T. Aoyama and J. Goryo. 
This work was partially supported by the special Grant-in-Aid for 
Promotion of Education and Science in Hokkaido University provided by the 
Ministry of Education, Science, Sport, and Culture, and by 
the Grant-in-Aid for 
Scientific Research on Priority area (Physics of CP violation) 
(Grant No. 12014201).

\begin{figure}
\epsfxsize=3in\centerline{\epsffile{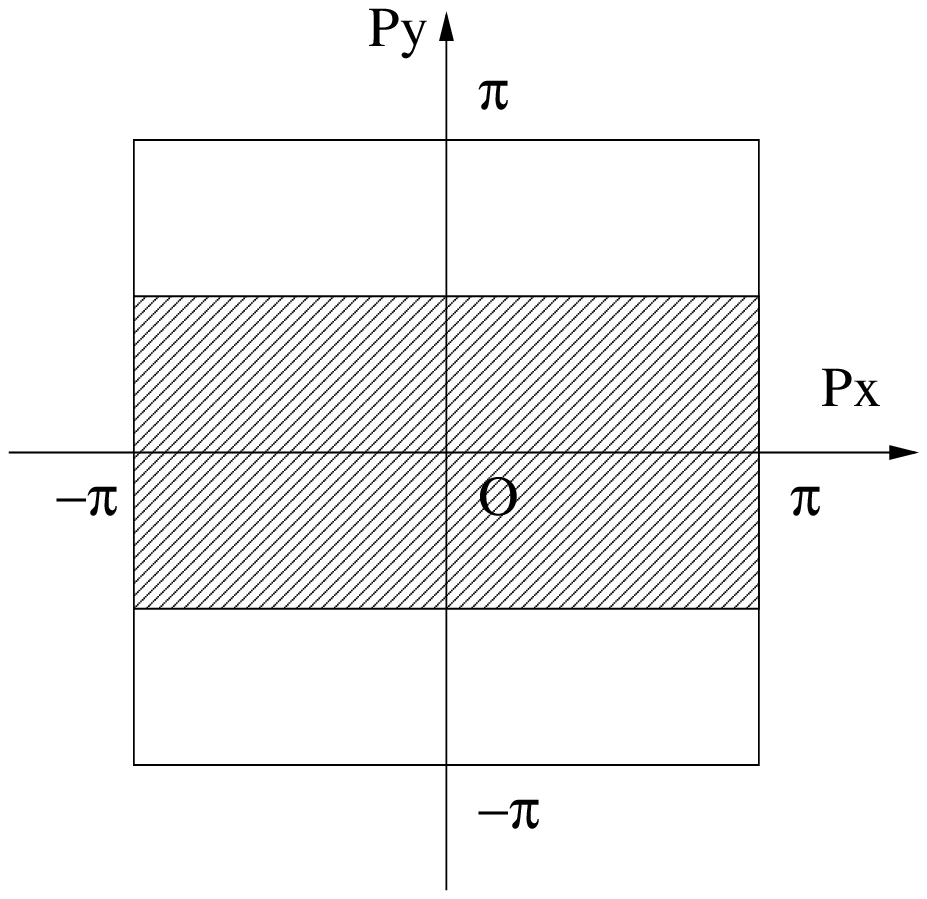}}
Fig.~1. The Fermi sea for the $K_x$ invariant Hall gas state. 
\end{figure}
\begin{figure}
\epsfxsize=3in\centerline{\epsffile{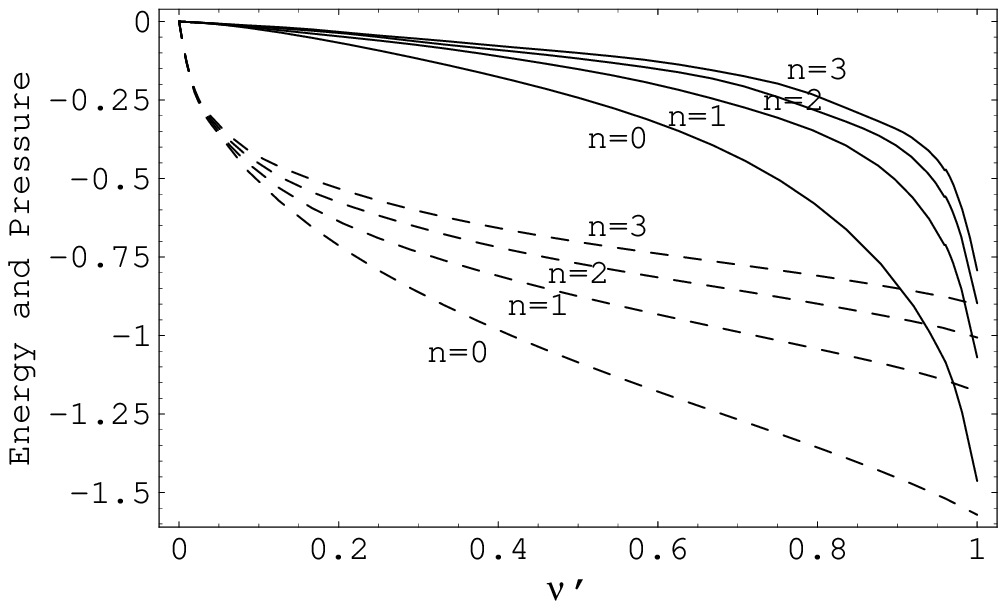}}
Fig.~2. Energy per particle (dashed lines) in the 
unit of $q^2/a$ and pressure (solid lines) in the unit of 
$q^2/a^3$ for $\nu=n+\nu'$, $n=$0, 1, 2, and 3. 
\end{figure}

\begin{figure}
\epsfxsize=3in\centerline{\epsffile{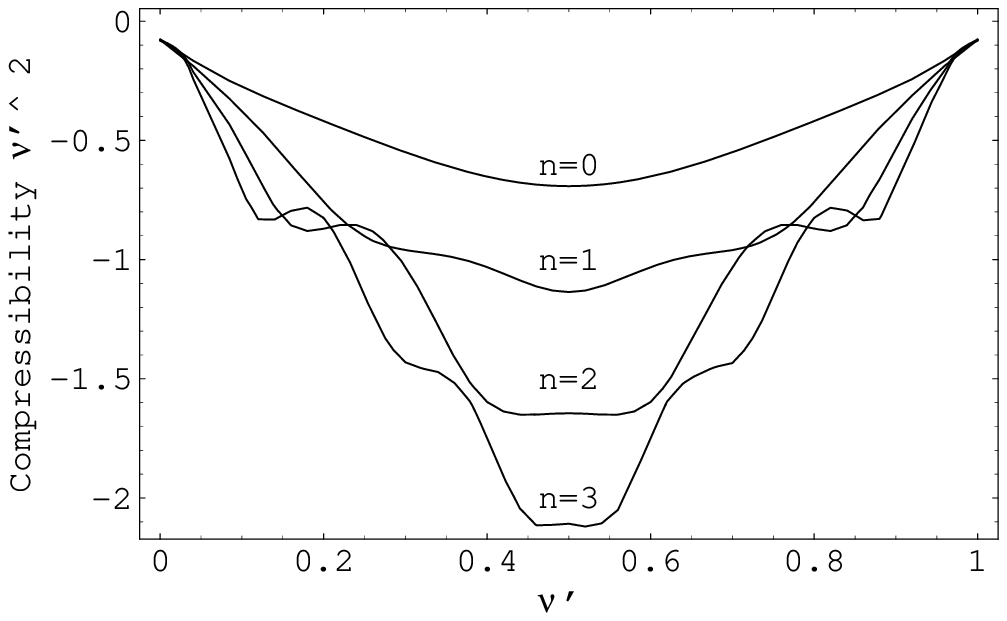}}
Fig.~3. Compressibility times $\nu'^2$ in the unit of $a^3/q^2$ 
for $\nu=n+\nu'$, $n=$0, 1, 2, and 3. 
\end{figure}

\begin{figure}
\epsfxsize=3in\centerline{\epsffile{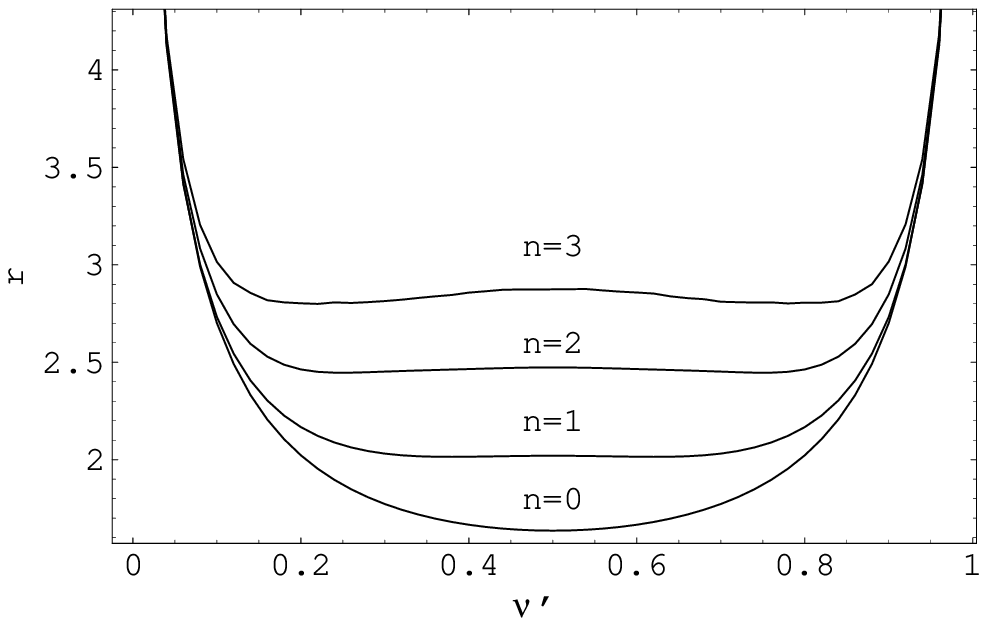}}
Fig.~4. Asymmetric parameter $r$ which minimizes the total energy for 
$\nu=n+\nu'$, $n=$0, 1, 2, and 3. 
\end{figure}

\begin{figure}
\epsfxsize=2in\centerline{\epsffile{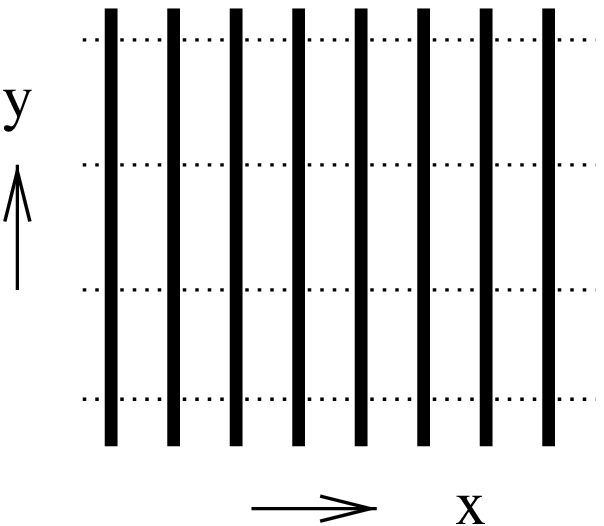}}
Fig.~5. Solid lines stand for the stripe ordering and dotted lines 
stand for the external potential. 
\end{figure}
\begin{figure}
\epsfxsize=3in\centerline{\epsffile{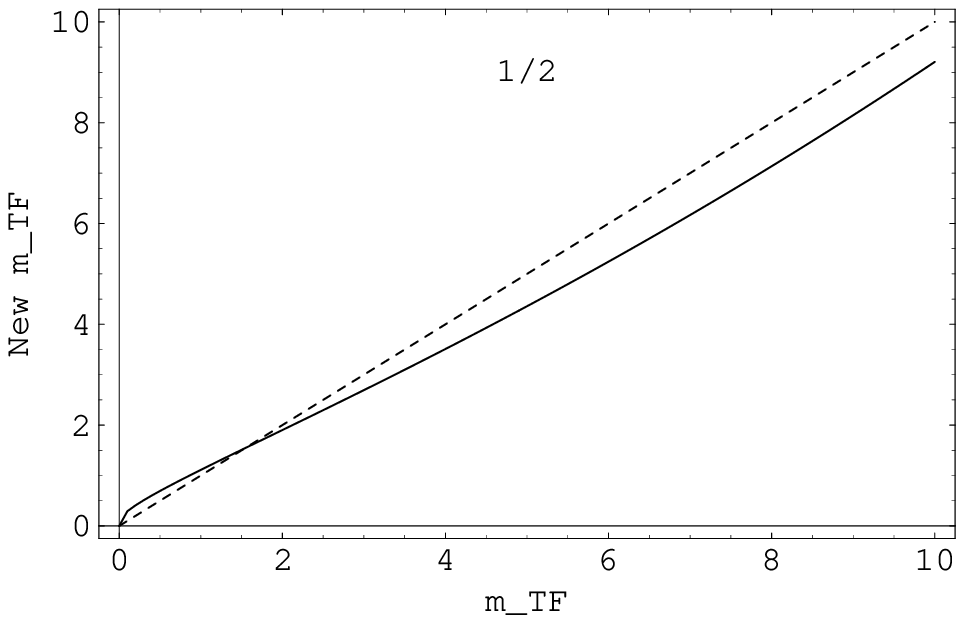}}
\epsfxsize=3in\centerline{\epsffile{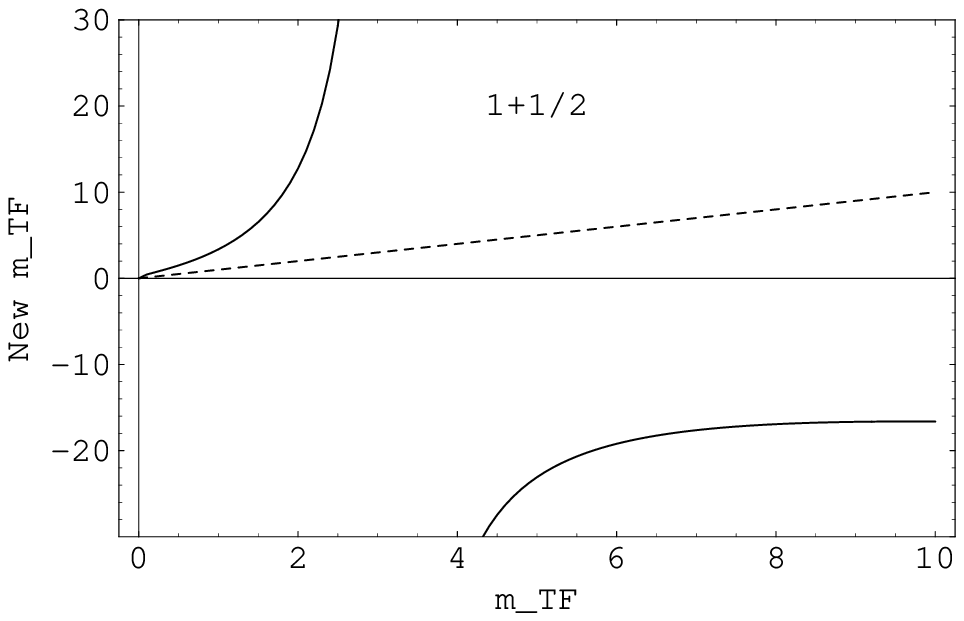}}
\epsfxsize=3in\centerline{\epsffile{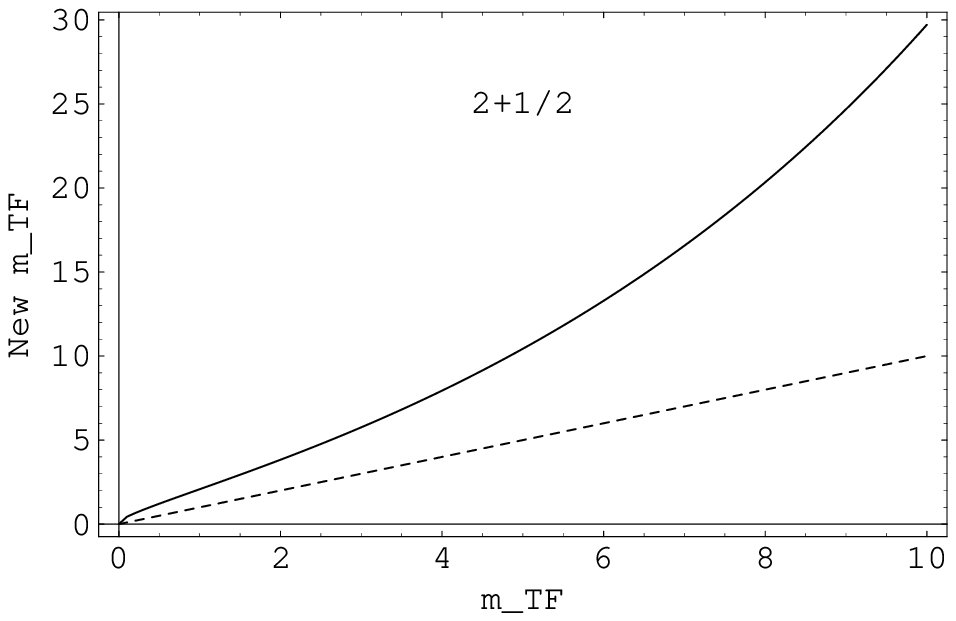}}
\epsfxsize=3in\centerline{\epsffile{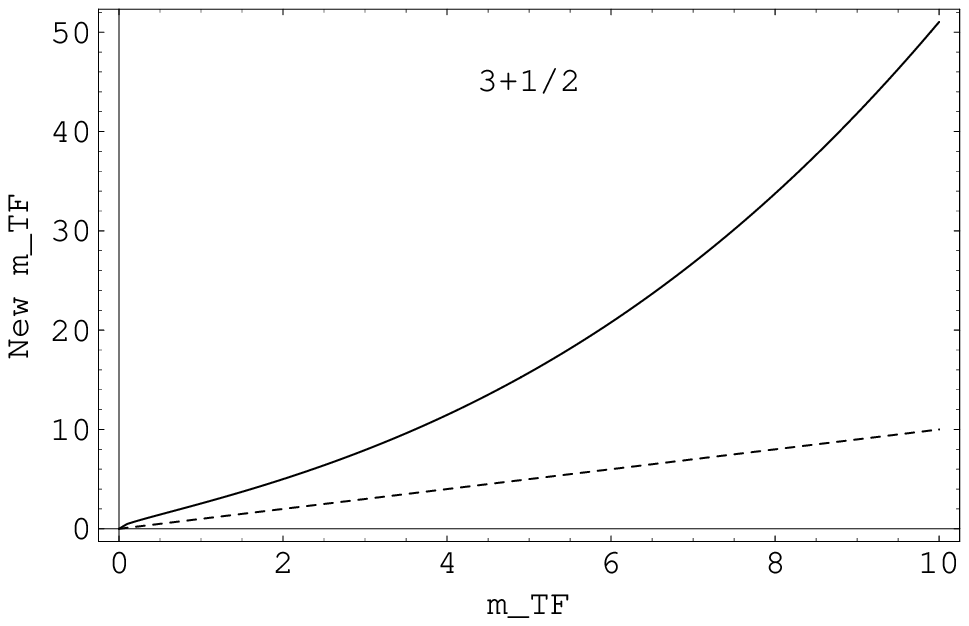}}
Fig.~6. Relation between initial TF screening mass and new TF screening mass 
for $\nu=$1/2, 1+1/2, 2+1/2, and 3+1/2. The asymmetric parameter is fixed 
at $r=2$. The dotted lines correspond to the points where initial TF 
screening mass and new TF screening mass are equivalent. 
\end{figure}
\begin{figure}
\epsfxsize=3in\centerline{\epsffile{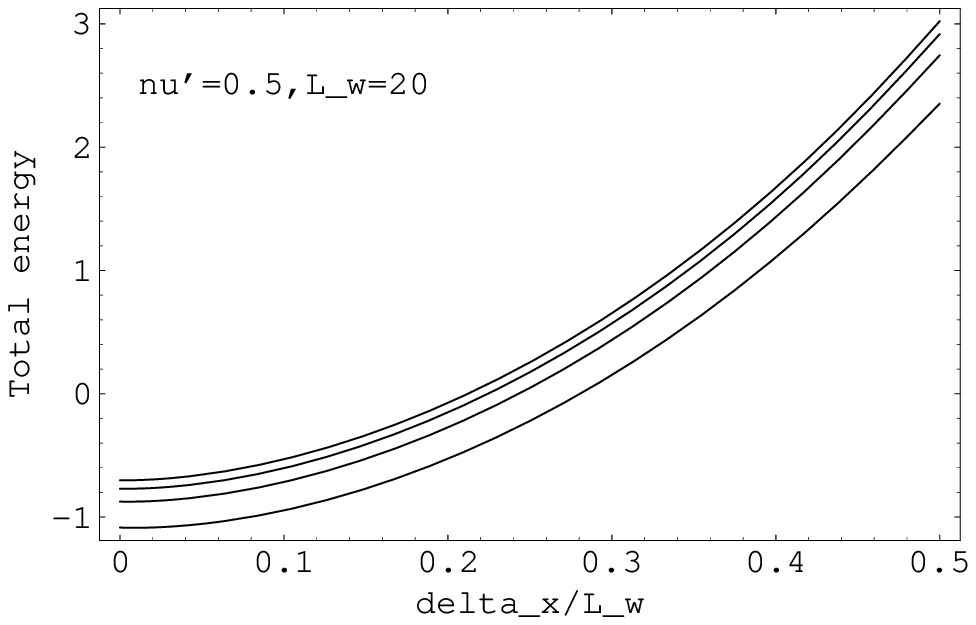}}
\epsfxsize=3in\centerline{\epsffile{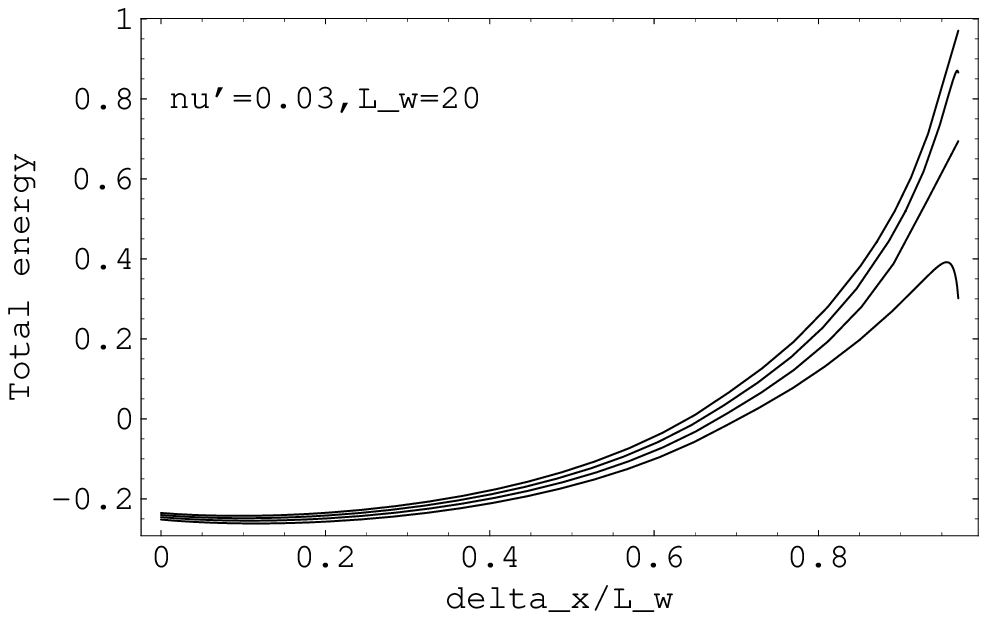}}
Fig.~7. The total energy versus $\delta_x/L_w$ for $\nu=n+\nu'$, 
$n=$0, 1, 2, and 3 (in order from below in both figures). 
The width of the strip is $L_w=20a$ and $\nu'=$0.03 and 0.5. 
The unit of the energy is $q^2/a$. 
\end{figure}
\begin{figure}
\epsfxsize=3in\centerline{\epsffile{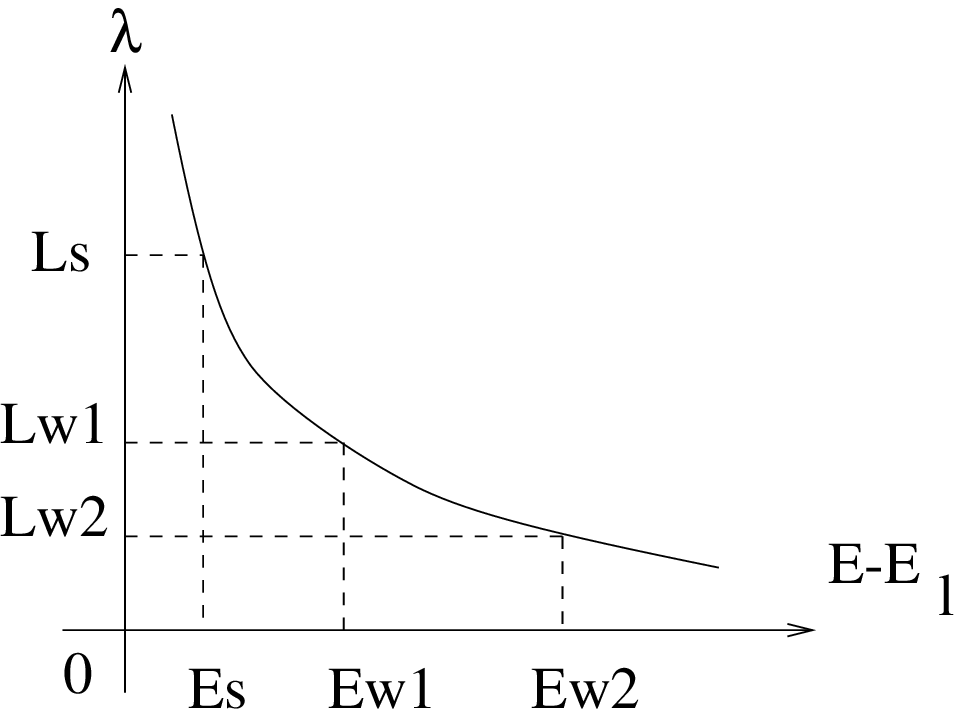}}
Fig. 8. Localization length versus the one-particle energy in the 
quantum Hall system. 
\end{figure}
\begin{figure}
\epsfxsize=3in\centerline{\epsffile{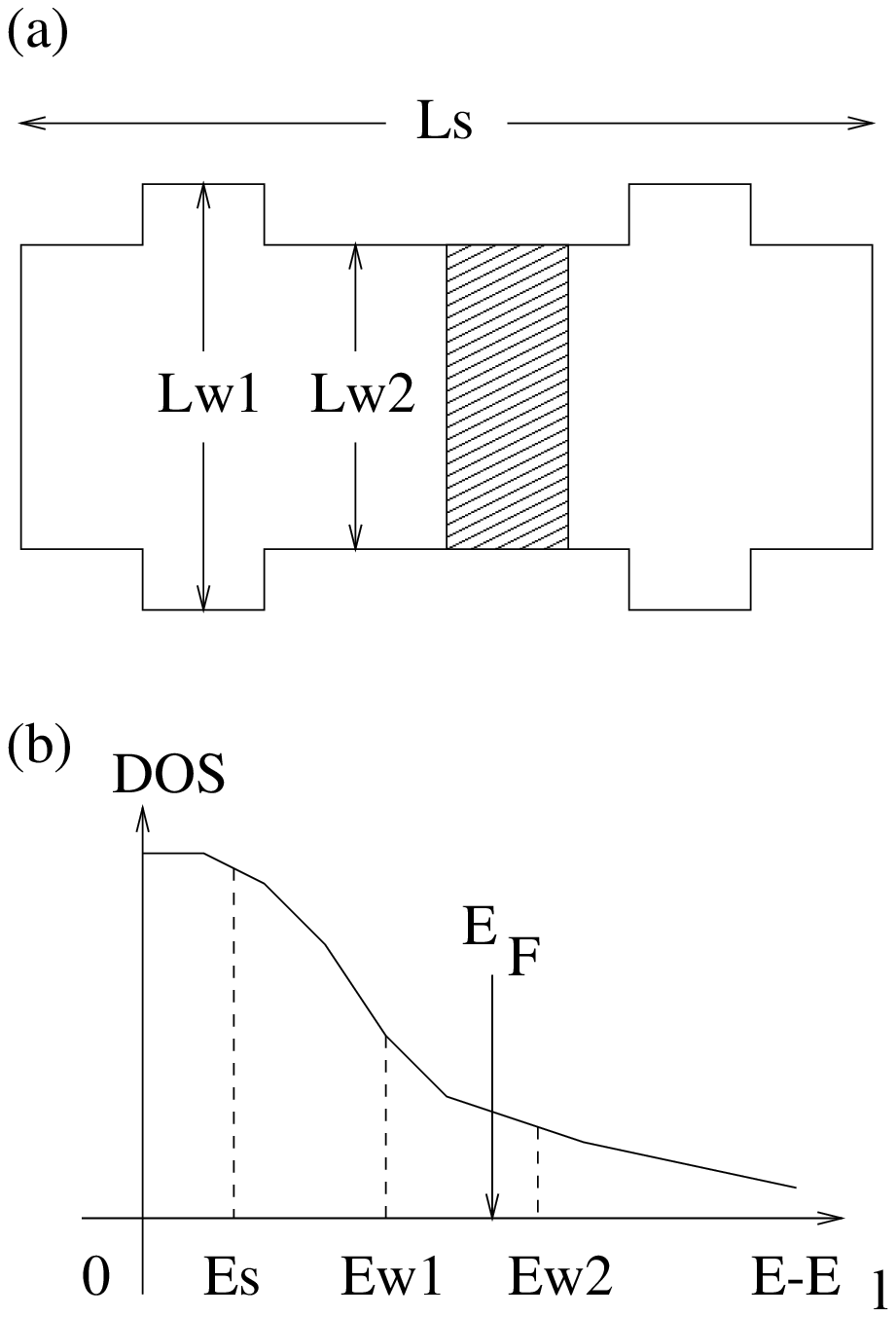}}
Fig. 9. A typical wave function in the configuration space (shaded region 
in (a)) and energy region in the density of states (DOS) in (b). 
The Fermi energy is in the dissipative QHR and the compressible state 
extends to the potential probe regions. 
\end{figure}
\begin{figure}
\epsfxsize=3in\centerline{\epsffile{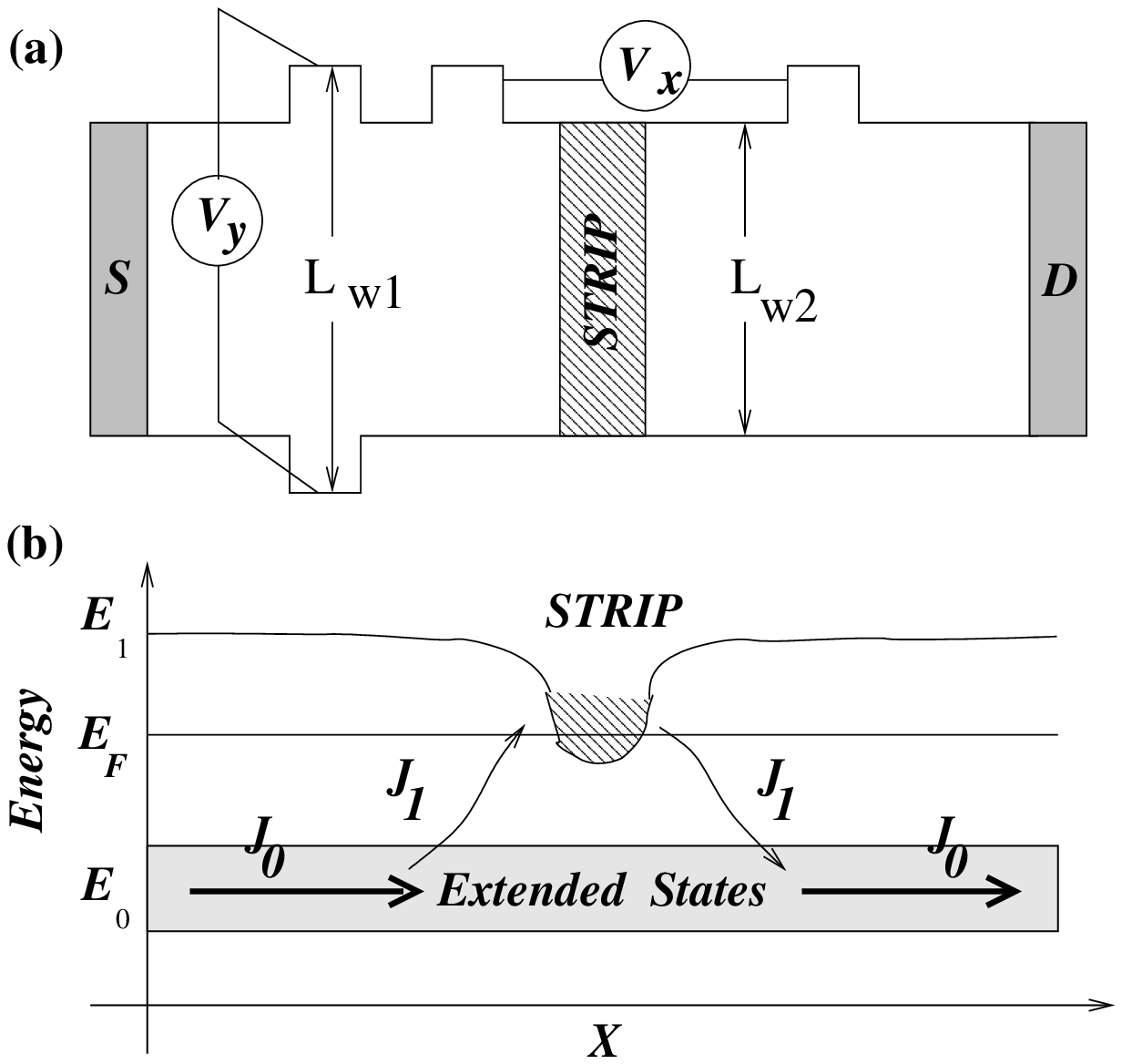}}
Fig~10. (a) Schematical view of a Hall bar in the dissipative QHR. 
Electric current is injected from S to D. 
(b) Sketch of the extended states carrying the undercurrent $J_0$ 
and the compressible state carrying the induced current $J_1$ at 
the strip region in the energy and the x-position.  
\end{figure}
\begin{figure}
\epsfxsize=2in\centerline{\epsffile{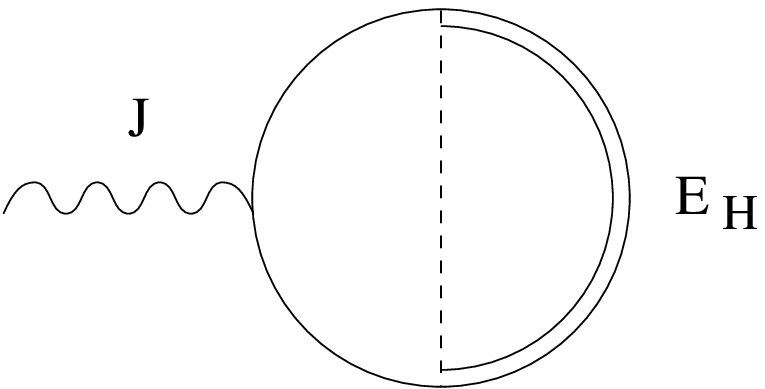}}
Fig. 11. Feynman diagram for the induced current $J$ activated from the 
undercurrent due to the Hall electric field $E_H$. 
\end{figure}
\begin{figure}
\epsfxsize=3in\centerline{\epsffile{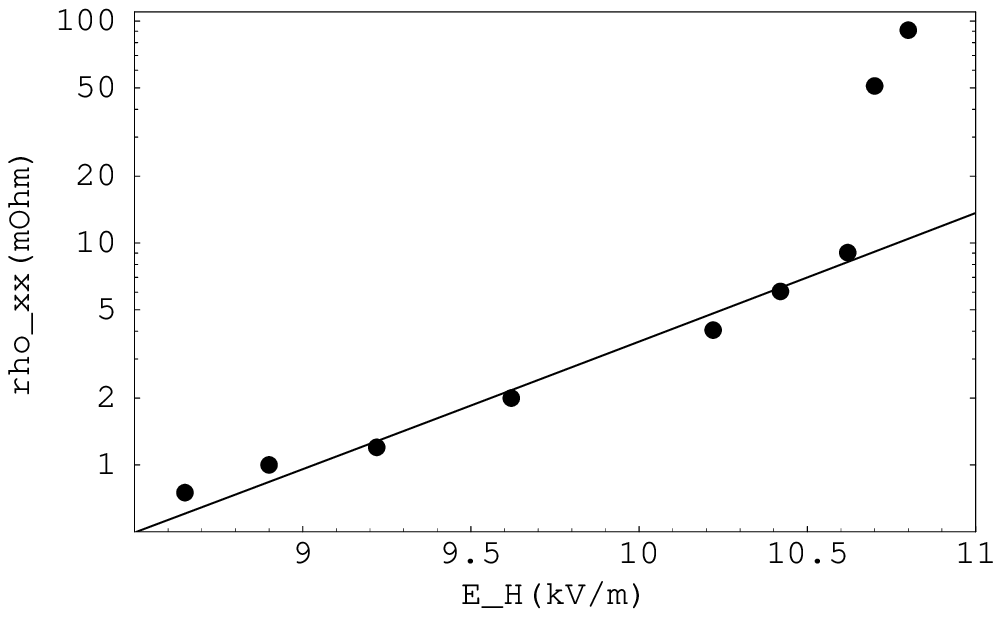}}
Fig. 12. Comparison with the experiment by Kawaji et al.\cite{b} for the 
Hall electric field dependence of longitudinal resistivity. 
Dots stand for experimental data and solid line is calculated by 
Eq.~(\ref{rho}). 
\end{figure}
\end{multicols}
\end{document}